\documentclass[a4paper,fleqn,final]{cas-dc}

\usepackage[T1]{fontenc}
\usepackage[utf8]{inputenc}
\usepackage[main=english]{babel}
\DeclareUnicodeCharacter{03C1}{\ensuremath{\rho}}
\DeclareUnicodeCharacter{03B2}{\ensuremath{\beta}}
\DeclareUnicodeCharacter{03BC}{\ensuremath{\mu}}
\DeclareUnicodeCharacter{03C4}{\ensuremath{\tau}}
\usepackage{textcomp}
\usepackage{hyperref}
\usepackage{amsmath,amssymb}
\usepackage{graphicx}
\usepackage{booktabs}
\usepackage{tabularx}
\usepackage{array}
\usepackage{calc}

\usepackage{threeparttable}
\usepackage{longtable}
\usepackage{adjustbox}
\usepackage[numbers,sort&compress]{natbib}
\usepackage{caption}
\usepackage{etoolbox}
\usepackage{placeins}
\usepackage{needspace}
\usepackage{stfloats}
\usepackage{xurl}
\usepackage{microtype}
\usepackage{lastpage}

\extrafloats{100}
\makeatletter
\g@addto@macro{\UrlBreaks}{\do\_\do\-\do\/\do\.\do\0\do\1\do\2\do\3\do\4\do\5\do\6\do\7\do\8\do\9}
\makeatother

\renewcommand{\ttfamily}{\rmfamily}
\AtBeginDocument{%
  \urlstyle{same}%
}

\usepackage{xcolor}
\definecolor{linknavy}{RGB}{0,70,127}
\hypersetup{
  colorlinks=true,
  linkcolor=black,
  citecolor=linknavy,
  urlcolor=linknavy,
  runcolor=linknavy
}

\AtBeginDocument{%
  \raggedbottom
  \setlength{\abovedisplayskip}{2pt plus 1pt minus 1pt}%
  \setlength{\belowdisplayskip}{2pt plus 1pt minus 1pt}%
  \setlength{\abovedisplayshortskip}{0pt}%
  \setlength{\belowdisplayshortskip}{2pt plus 1pt minus 1pt}%
}

\AtBeginEnvironment{table}{\let\sffamily\rmfamily}
\AtBeginEnvironment{figure}{\let\sffamily\rmfamily}
\AtBeginEnvironment{table*}{\let\sffamily\rmfamily}
\AtBeginEnvironment{figure*}{\let\sffamily\rmfamily}

\ExplSyntaxOn
\RenewDocumentCommand \firstname {}
  { \textcolor{black}{\seq_use:Nn \l_stm_au_seq { ~ }} }

\RenewDocumentCommand \emailauthor { m m }
   {
     \int_gincr:N \g_ead_int
     \seq_gput_right:Nn \g_stm_ead_seq
       {
         { \href{mailto:#1}{\rmfamily #1} }
         \parsename { #2 }
         \space(\eadauthor)
       }
     }

\cs_set:Npn \__first_footerline:
{
  \group_begin:
  \small
  \normalfont
  \ifnum\theblind>0\relax
  \else
  \__short_authors: :~
  \fi
  \itshape Preprint~submitted~to~Frontiers
  \group_end:
}

\RenewDocumentCommand \printorcid { } { }

\cs_set:Npn \__first_head:
{
  \parbox[t]{\textwidth}
  {
    \rule{\textwidth}{0pt}
  }
}

\cs_set:Npn \__cas_head:
{
  \parbox{\textwidth}
  {
    \rule{\textwidth}{0pt}
  }
}

\cs_set:Npn \__cas_foot:
{
  \parbox[t]{\textwidth}
  {
   \rule{\textwidth}{.2pt}\\
   \small
   \normalfont
   \__first_footerline:
   \hfill Page~\thepage {}~of~ \lastpage
  }
}
\ExplSyntaxOff

\makeatletter
\ps@cas
\makeatother

\begin{document}

\makeatletter
\def\bstctlcite{\@ifnextchar[{\@bstctlcite}{\@bstctlcite[@auxout]}}
\def\@bstctlcite[#1]#2{\@bsphack
  \@for\@citeb:=#2\do{%
    \edef\@citeb{\expandafter\@firstofone\@citeb}%
    \if@filesw\immediate\write\csname #1\endcsname{\string\citation{\@citeb}}\fi}%
  \@esphack}
\makeatother
\bstctlcite{IEEEbsTcontrol}

\shortauthors{Farajpoor et al.}
\shorttitle{Vineyard wildfire resilience}

\title[mode=title]{Multisource Remote Sensing and Geospatial Analysis of Vineyard Wildfire Impacts and Exposure: The 2019 Kincade Fire}

\author[1]{Parastoo Farajpoor}
\author[2]{Mahla Ardebili Pour}
\author[3]{Mohammad Bagher Ghiasi}
\author[1]{Mohammadreza Narimani}
\cormark[1]
\ead{mnarimani@ucdavis.edu}

\affiliation[1]{organization={Department of Biological and Agricultural Engineering, University of California, Davis},city={Davis},state={CA},postcode={95616},country={USA}}
\affiliation[2]{organization={Department of Civil and Environmental Engineering, University of California, Davis},city={Davis},state={CA},postcode={95616},country={USA}}
\affiliation[3]{organization={Department of Electrical and Computer Engineering, University of California, Davis},city={Davis},state={CA},postcode={95616},country={USA}}

\cortext[1]{Corresponding author}

\begin{abstract}
Working agricultural landscapes are often treated as background to wildfire disasters, even though they are managed fuel mosaics, productive assets, and parts of regional infrastructure systems. We examine vineyard wildfire resilience during the electrically initiated 2019 Kincade Fire in Sonoma County, California, using an open, event-anchored geospatial framework spanning 4,581 vineyard fields (8,813.2~ha), wildland vegetation, surveyed structures, roads, overhead smoke, and post-fire greenness. Sentinel-2 spectral response, OpenET evapotranspiration, gridMET fire weather, USDA soils, USGS terrain, NOAA smoke polygons, an ignition-date OpenStreetMap network, and three-dimensional data inventories were analyzed at their native decision scales. Vineyard pixels displayed substantially lower descriptive dNBR than wildland pixels inside the perimeter (means 0.130 and 0.337, respectively). Yet this landscape contrast did not identify a universal vineyard firebreak effect. A segment-clustered boundary model yielded a small negative contrast at 100~m ($\tau = -0.0166$) but changed across bandwidths, failed slope continuity, disappeared in a 100~m donut specification, and produced a large wrong-signed placebo estimate. A 250~m spatial generalized additive model also reversed the unconditional pattern: after conditioning on location, terrain, and water use, vineyard fraction was positively associated with dNBR, while residual Moran's~$I$ remained 0.519. Beyond direct spectral impact, all mapped vineyards intersected overhead smoke on at least one day (mean 7.78 potential smoke-days per field), 34.2\% of road-network nodes were dead ends, and vineyards within the perimeter showed a larger greenness deficit through 2021 (recovery ratios 0.815 inside and 0.854 outside). Lower immediate spectral impact therefore did not imply complete resilience. The study offers a reproducible urban-rural informatics template that separates descriptive contrasts, conditional associations, exposure indicators, and recovery evidence for decision-making in working landscapes.
\end{abstract}


\maketitle

\section{Introduction}\label{introduction}

Wildfire risk in the western United States is commonly framed through the wildland-urban interface, where housing expansion, vegetation, and extreme fire weather combine to expose communities. That framing remains essential, but it does not fully represent working agricultural landscapes that occupy the same fire-prone regions. Human ignitions now extend the fire niche across seasons and locations, anthropogenic warming has increased fuel aridity, and the rapidly expanding interface between built and vegetated land has raised both exposure and loss \citep{abatzoglou2016climate,balch2017human,radeloff2018wui}. In California's North Coast, vineyards, rural communities, electrical corridors, roads, woodland, chaparral, and grassland form a coupled landscape in which fire consequences move between production, infrastructure, ecosystems, and people.

This agricultural interface is not simply a variation of the conventional WUI. A vineyard is a managed agroecosystem whose water use, trained canopy, inter-row cover, soil setting, and access network differ from neighboring wildland fuels. Its exposure is also multidimensional: vines may experience direct combustion, smoke during harvest, disrupted access, and delayed post-fire recovery even when spectral burn response is lower than in surrounding vegetation. Coupled human-natural systems and cascading-hazard scholarship provide a useful conceptual basis because they ask how disturbances propagate across interdependent subsystems rather than treating each map layer as an independent risk score \citep{chen2014cnh,moftakhari2019compound,alcantara2025cascading,li2025coupled}. Cross-hazard urban-resilience research has likewise used machine-learning tools to integrate multidimensional indicators over time, illustrating the value of data-driven decision support in resilience planning \citep{pour2025ml}.

Remote sensing makes these interactions observable, but interpretation over vineyards requires care. Spectral indices can reveal canopy moisture, greenness, and change; thermal imagery can indicate water stress; and field-scale evapotranspiration products provide a physically interpretable measure of crop water use. Vineyard studies have demonstrated the value and uncertainty of thermal and multispectral water-status retrievals and satellite-based ET at field scale \citep{bellvert2015cwsi,kalua2020vineyardet}. Recent grapevine spectral research also shows that canopy and leaf spectra encode multiple physiological and biochemical traits, reinforcing the need to interpret spectral change as a mixture of canopy amount, condition, and background rather than a direct measure of combustion alone \citep{farajpoor2025grapevine}. Field-scale Sentinel-2 applications are increasingly useful for agricultural decision support, but cloud gaps, ground truth, and transferability remain persistent limitations \citep{narimani2026c}.

The Kincade Fire offers a particularly informative case. It began near the Geysers geothermal field on 23 October 2019, burned 77,758 acres according to the final incident record, destroyed 374 structures, and prompted the largest evacuation in Sonoma County history before containment on 6 November \citep{sonoma2020kincadeaar}. Related Sonoma County modeling work has examined wildfire growth on heterogeneous landscapes for prevention planning \citep{sayarshad2025sonoma}. Its initiating mechanism is not inferred in this study: CAL FIRE attributed the fire to Pacific Gas and Electric transmission infrastructure, and the California Public Utilities Commission documented the associated safety violations and administrative resolution \citep{calfire2020kincade,cpuc2021consentorder,cpuc2021kincadeinvestigation}. The event also unfolded during concurrent public-safety power shutoffs and severe wind and humidity conditions, making it a clear example of infrastructure and weather acting within an agricultural-community system \citep{brown2022sonoma,purdy2020psps}.

The necessary methods are individually established. Monitoring Trends in Burn Severity formalized consistent satellite assessment of fire effects, while NBR, dNBR, and RdNBR remain standard spectral indicators with known ecological and sampling constraints \citep{eidenshink2007mtbs,key2006severity,miller2007rdnbr}. Sentinel-2 and Google Earth Engine allow these products to be reproduced at landscape scale \citep{drusch2012sentinel2,gorelick2017gee}, and OpenET provides operational field-scale ET estimates for the western United States \citep{melton2022openet}. Three-dimensional structure can be described with airborne lidar, GEDI, and ICESat-2, although sparse sampling, temporal mismatch, and scale remain important limitations for vineyard parcels \citep{dubayah2020gedi,neuenschwander2019atl08,neumann2019icesat2,malambo2024canopy}. Lidar-derived canopy fuel attributes can inform fire-behavior modeling, but reliable event-day interpretation requires acquisition dates that actually represent the fire period \citep{engelstad2019canopyfuel,pascual2026firebehavior}.

A second challenge is inference. Large pixel samples can make almost any landscape contrast appear statistically decisive even when neighboring pixels are not independent. Geographic boundary designs can sharpen comparisons, but only when potential confounders are locally continuous and placebo tests are credible \citep{lee2010rdd,calonico2014robust,keele2015geographic,wuepper2020spatialrdd}. Similarly, predictive-model accuracy or flexible spatial smooths cannot substitute for spatially honest diagnostics. Recent data-driven wildfire modeling illustrates the value of comparative approaches, yet also shows that greater algorithmic complexity does not necessarily yield more transferable conclusions \citep{biswas2025geoai}. The same lesson appears in spatially validated urban wildfire analysis, where random validation can materially overstate generalization \citep{farajpoor2026palisades}.

The remaining consequences extend beyond the burn scar. Wildfire smoke has well-established health effects, but an overhead satellite smoke polygon is neither a ground-level PM$_{2.5}$ measurement nor evidence of grape smoke taint, which requires chemical or sensory assessment \citep{reid2016smokehealth,krstic2015smoketaint,summerson2021smokereview}. California wine-grape growers nevertheless report substantial wildfire and smoke-management uncertainty \citep{zakowski2023winegrape}. Transportation studies likewise show that evacuation behavior and network conditions shape community outcomes, but road topology alone cannot reveal realized traffic performance \citep{wong2023evacuee,borody2025evacuation,xu2023kincade}. Post-fire soil and hydrologic change and documented ecological responses to Kincade add further dimensions of recovery \citep{neris2023soilerosion,usgs2025postfiresoils,lumpkin2026birds}.

Resilience describes how well a system can endure a disturbance without losing its core functions. It also reflects the system's ability to adjust and recover after the event \citep{meerow2016urbanresilience,ardebili2024floodresilience}. In the context of this study, vineyard wildfire resilience is therefore considered a multidimensional property of the managed agricultural landscape, encompassing conditions that may influence resistance to wildfire impacts, the magnitude of direct and indirect exposure during the event, and the capacity for post-fire recovery.
Accordingly, we evaluate resilience through complementary indicators of pre-fire vegetation and water status, direct spectral fire impact, smoke exposure, transportation accessibility, and post-fire vegetation recovery. These indicators represent distinct dimensions of resilience rather than components of a single composite resilience score.

Here we ask a bounded question: what can openly available geospatial evidence establish about immediate spectral impact, vineyard water status, spatial boundary behavior, transportation access, smoke exposure, structural-data readiness, and short-term recovery during one well-documented agricultural-interface wildfire? We use all subsystems to tell a single story, but we do not collapse them into an arbitrary resilience score. Instead, each result is tied to a stated analytical unit and claim boundary (Table~\ref{tab:claims}). The study contributes an event-level assessment of vineyard-wildland contrasts and a reproducible urban-rural informatics framework in which lower immediate spectral impact, persistent exposure, and recovery deficits can coexist. This form of transparent screening complements, rather than replaces, field observations and operational inventories \citep{narimani2026b}.

\section{Materials and Methods}\label{materials-and-methods}

\subsection{Study design and analytical hierarchy}\label{study-design}

The study area comprises the 2019 Kincade Fire footprint and its immediate surroundings in northern Sonoma County, centered near $122.78^{\circ}$~W and $38.79^{\circ}$~N. Vineyards occupy valley floors and lower slopes of the Alexander and Russian River valleys, while oak woodland, chaparral, and annual grassland dominate steeper terrain. All distance and area calculations used NAD83 / UTM zone 10N (EPSG:26910).

Two nested areas of interest were used. A 5~km perimeter buffer supported landscape context, smoke, roads, recovery, and the complete vineyard inventory (4,581 fields; 8,813.2~ha). A perimeter-plus-2~km analysis area supported severity and boundary comparisons (2,622 vineyard fields; 5,798.3~ha). The official incident total was 77,758 acres. The WFIGS/NIFC polygon carried a GIS\_ACRES attribute of 77,762 acres; small differences produced by geometry reprojection were treated as cartographic rather than scientific results.

\begin{figure*}[!t]
\centering
\includegraphics[width=\textwidth]{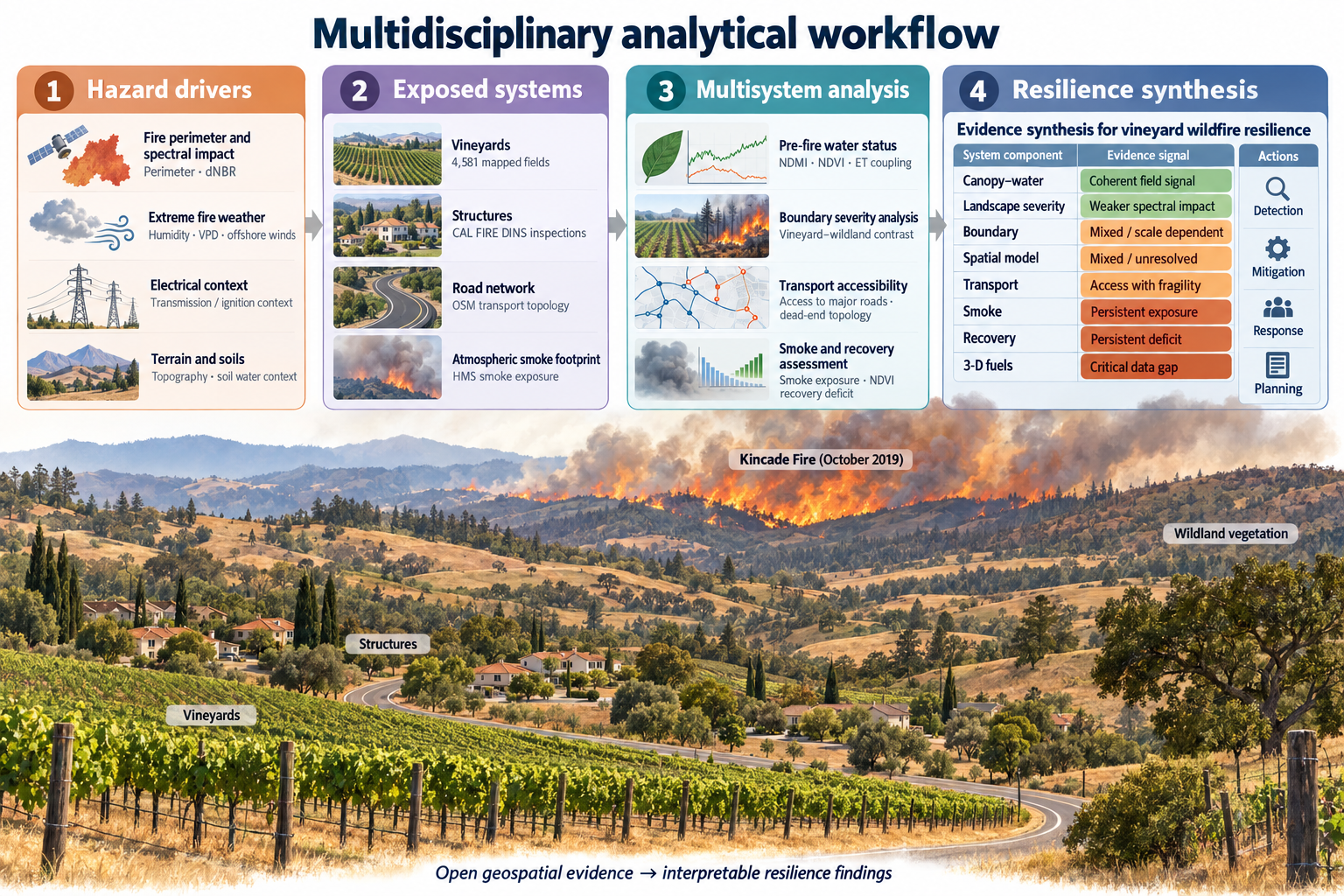}
\caption{Multidisciplinary analytical workflow. Four-stage evidence pathway from hazard drivers (fire extent and spectral response, fire weather, electrical context, terrain, and soils) through exposed systems (vineyards, surveyed structures, the 2019 road network, and overhead smoke), multisystem analyses (pre-fire water status, vineyard-wildland boundary contrasts, transport accessibility, smoke exposure, and recovery), and a qualitative evidence synthesis. The landscape scene is a conceptual illustration, not event imagery from the 2019 Kincade Fire.}
\label{fig:workflow}
\end{figure*}

The evidence pathway is summarized in Figure~\ref{fig:workflow} and the study geography in Figure~\ref{fig:studyarea}. Variables were classified before analysis as baseline conditions, event or impact indicators, or recovery indicators. Post-fire variables were never used to explain pre-fire state. Table~\ref{tab:datasets} lists the principal public datasets, and Table~\ref{tab:claims} links each subsystem to its analytical unit, method, and claim boundary.

\subsection{Fire, vineyard, structure, and environmental data}\label{data}

The final perimeter was obtained from the WFIGS/NIFC InterAgency Fire Perimeter History and matched by incident name. CAL FIRE Damage Inspection data provided 1,568 surveyed structures: 374 destroyed, 6 with major damage, 14 with minor damage, 40 affected, and 1,134 with no recorded damage. Structure points were retained as contextual direct-impact observations rather than a modeled outcome.

\begin{figure*}[!t]
\centering
\includegraphics[width=\textwidth]{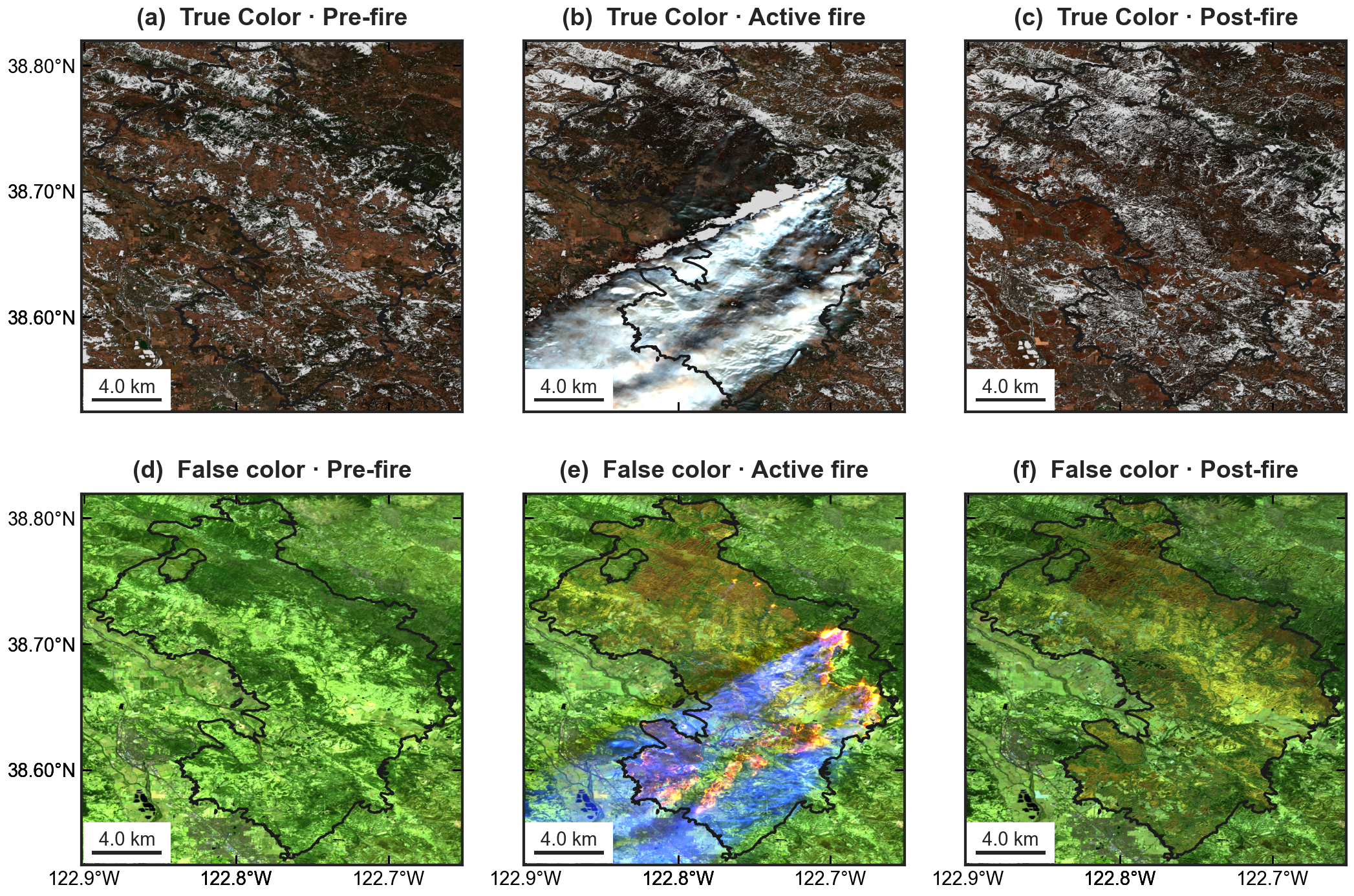}
\caption{Sentinel-2 event chronology in true and shortwave-infrared color. (A--C) Sentinel-2 true-color mosaics for pre-fire, active-fire (27 October 2019), and first-clean post-fire conditions. (D--F) Shortwave-infrared false-color views for the same dates, emphasizing active fire and the post-fire spectral footprint. Active-burn imagery is shown for context only and was not used to calculate the pre-post severity response.}
\label{fig:chronology}
\end{figure*}

Vineyard fields were extracted from the California Department of Water Resources 2019 Statewide Crop Mapping geodatabase using the vineyard class and a stable field identifier. These polygons define the spatial support of the measurements, rather than land ownership or a new image-derived segmentation. This distinction is also explicit in recent NAIP-based farmland mapping, which separates visible agricultural extent from cadastral parcels \citep{narimani2026a}. Sentinel-2 SR Harmonized imagery supplied optical reflectance; Landsat 8 Collection 2 Level 2 supplied an independent landscape-scale spectral check. OpenET ensemble monthly ET v2.1 represented April--October 2019 crop water use. gridMET supplied daily relative humidity, vapor-pressure deficit, and wind speed for a contextual window from 10 October through 6 November 2019 (including pre-ignition lead-in days; ignition marked on 23 October) \citep{abatzoglou2013gridmet}. USDA NRCS SSURGO supplied available water capacity (AWC), and USGS 3DEP supplied elevation, slope, northness, and eastness. NOAA/NESDIS Hazard Mapping System polygons represented daily potential overhead smoke, and a date-fixed OpenStreetMap drive network represented road geometry on 23 October 2019. Figure~\ref{fig:chronology} shows the optical event chronology.

\subsection{Spectral indices and burn-response products}\label{spectral-indices}

Cloud- and shadow-screened Sentinel-2 median composites were created for 1 September--22 October 2019 (71 usable scenes) and 10 November--15 December 2019 (24 scenes). Active-fire imagery from 27 October was retained only for visual chronology. All calculations were implemented in Google Earth Engine with collection identifiers, filters, masks, scales, and export parameters recorded in the repository.

\clearpage
\noindent
\begin{minipage}{\columnwidth}
\centering
\includegraphics[width=\columnwidth]{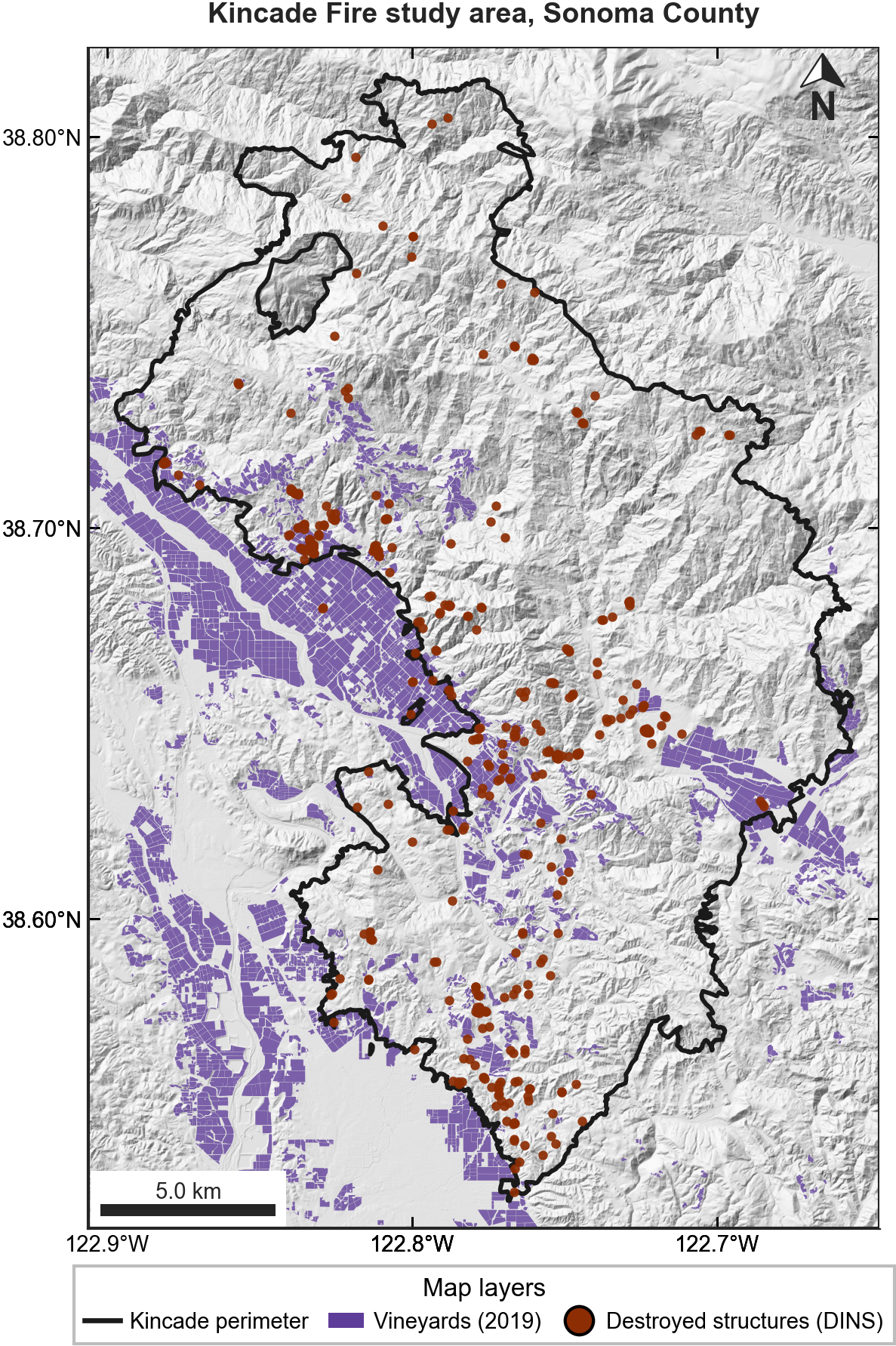}
\captionof{figure}{Kincade Fire study area. Fire perimeter (black outline), 2019 vineyard fields, and CAL FIRE Damage Inspection (DINS) points showing destroyed structures (dark ochre) and other surveyed structures (grey) over a USGS 3DEP hillshade. Vineyards are concentrated on valley floors and lower slopes along the western and southern portions of the burned landscape. The official incident total was 77,758 acres; the WFIGS polygon used in analysis carries a GIS\_ACRES value of 77,762 acres.}
\label{fig:studyarea}
\end{minipage}

\begin{equation}
\mathrm{NDVI}=\frac{\rho_{\mathrm{NIR}}-\rho_{\mathrm{red}}}{\rho_{\mathrm{NIR}}+\rho_{\mathrm{red}}}
\label{eq:ndvi}
\end{equation}
where NDVI is the Normalized Difference Vegetation Index; $\rho$ denotes surface reflectance, and the NIR and red subscripts denote the near-infrared and red bands, respectively.
\begin{equation}
\mathrm{NDMI}=\frac{\rho_{\mathrm{NIR}}-\rho_{\mathrm{SWIR1}}}{\rho_{\mathrm{NIR}}+\rho_{\mathrm{SWIR1}}}
\label{eq:ndmi}
\end{equation}
where NDMI is the Normalized Difference Moisture Index and the SWIR1 subscript denotes the first shortwave-infrared band; $\rho$ and the NIR subscript are defined as in Equation~\eqref{eq:ndvi}.
\begin{equation}
\mathrm{NBR}_{t}=\frac{\rho_{\mathrm{NIR},t}-\rho_{\mathrm{SWIR2},t}}{\rho_{\mathrm{NIR},t}+\rho_{\mathrm{SWIR2},t}}
\label{eq:nbr}
\end{equation}
where NBR is the Normalized Burn Ratio, $t$ indexes the acquisition period, and the SWIR2 subscript denotes the second shortwave-infrared band.
\begin{equation}
\mathrm{dNBR}=\mathrm{NBR}_{\mathrm{pre}}-\mathrm{NBR}_{\mathrm{post}}
\label{eq:dnbr}
\end{equation}
where dNBR is the differenced Normalized Burn Ratio and the pre and post subscripts denote the pre-fire and post-fire NBR composites, respectively.
\begin{equation}
\mathrm{RdNBR}=\frac{\mathrm{dNBR}}{\sqrt{\lvert\mathrm{NBR}_{\mathrm{pre}}\rvert}}
\label{eq:rdnbr}
\end{equation}
where RdNBR is the relativized differenced Normalized Burn Ratio, which scales dNBR by the magnitude of the pre-fire NBR.

\begin{table*}[!t]
\centering
\caption{Public datasets, spatial-temporal support, analytical roles, and principal limitations.}
\label{tab:datasets}
\footnotesize
\begin{tabular}{@{}p{0.13\textwidth}p{0.22\textwidth}p{0.14\textwidth}p{0.12\textwidth}p{0.30\textwidth}@{}}
\toprule
Domain / dataset & Source or identifier & Resolution / unit & Period & Analytical role and principal limitation \\
\midrule
Fire perimeter & WFIGS/NIFC InterAgency Fire Perimeter History & Incident polygon & 2019 & Study boundary; attribute 77,762 acres, official incident total 77,758 acres. \\
Structure damage & CAL FIRE POSTFIRE\_MASTER\_DATA\_SHARE / DINS & Point inspection & Post-fire 2019 & Contextual direct impact; not modeled as a structure-loss outcome. \\
Vineyards & California DWR 2019 Statewide Crop Mapping, vineyard class & Field polygon & 2019 & Field inventory and land-cover boundary; crop class does not encode cultivar or management. \\
Optical response & Sentinel-2 SR Harmonized & 10--20~m & 2018--2021 & NDVI, NDMI, NBR, dNBR, recovery; spectral response is not field-calibrated vine damage. \\
Cross-sensor check & Landsat 8 Collection 2 Level 2 & 30~m & 2019 & Landscape-average dNBR magnitude only; does not validate the vineyard-wildland contrast. \\
Water use & OpenET ensemble, gridMET monthly v2.1 & 30~m monthly & Apr--Oct 2019 & Field ET and recovery context; ET is not applied irrigation. \\
Fire weather & gridMET & $\sim$4~km daily & 10 Oct--6 Nov 2019 & Area-mean RH, VPD, and wind context (pre-ignition lead-in included); insufficient for row-scale spread. \\
Terrain and soils & USGS 3DEP; USDA NRCS SSURGO & 10~m raster; map unit & Static & Elevation, slope, aspect, AWC; map-unit soils simplify field variability. \\
Smoke & NOAA/NESDIS HMS smoke polygons & Daily polygon & 23 Oct--10 Nov 2019 & Potential overhead smoke; not ground PM$_{2.5}$ or smoke taint. \\
Road network & OpenStreetMap historical snapshot & Vector graph & 23 Oct 2019 & Access and topology; no volume, capacity, closure, or evacuation behavior. \\
3-D structure & OpenTopography, USGS 3DEP LPC, GEDI L2A, ICESat-2 inventory & Point cloud / footprint & 2003--2024 & Structural feasibility; no dedicated 2019 airborne collect and no parcel-scale event fuels. \\
\bottomrule
\end{tabular}
\end{table*}

\begin{table*}[!t]
\centering
\caption{Research questions, analytical units, methods, and explicit claim boundaries.}
\label{tab:claims}
\footnotesize
\begin{tabular}{@{}p{0.14\textwidth}p{0.20\textwidth}p{0.14\textwidth}p{0.20\textwidth}p{0.24\textwidth}@{}}
\toprule
Subsystem & Question & Analytical unit & Method / statistic & Claim boundary \\
\midrule
Pre-fire canopy condition & Does vineyard canopy condition track seasonal water use? & Field ($n{=}4{,}581$) & Pearson/Spearman association; descriptive OLS & Association only; ET is not metered irrigation. \\
Landscape spectral response & How different is dNBR over vineyards and wildland? & 20~m pixels & Means, medians, relative difference & Descriptive; pixel independence not assumed. \\
Boundary behavior & Does dNBR change near vineyard-wildland edges? & 2,000 boundary segments & Local-linear model with clustered SE; bandwidth, continuity, donut, placebo & Spatial boundary association, not causal RDD. \\
Conditional landscape pattern & Which measured covariates accompany dNBR? & 250~m cells ($n{=}5{,}097$) & Penalized spatial GAM; Moran residual diagnostic & Conditional diagnostic; residual autocorrelation remains. \\
Within-vineyard mechanism & Does field ET covary with field dNBR? & Fields with complete covariates ($n{=}1{,}206$) & Standardized OLS & Spectral/biomass association; no irrigation-effect claim. \\
Smoke exposure & How often were vineyards beneath mapped smoke? & Field-day & HMS intersection and hectare-smoke-days & Potential overhead exposure; not dose or taint. \\
Road access & How are vineyards situated in the drive network? & Field centroid / graph & Nearest-distance and network topology metrics & Geometry/topology only; not evacuation performance. \\
Recovery & Did inside-perimeter vineyard greenness diverge? & Fields (1,327 inside; 3,254 outside) & Season-matched NDVI ratios; unadjusted group tests & Greenness only; observational, not yield or causal fire effect. \\
3-D structure & Were event-day vertical fuels observable? & Dataset / footprint inventory & Coverage and vintage audit & Data-readiness conclusion only. \\
\bottomrule
\end{tabular}
\end{table*}

\subsection{Pre-fire vineyard canopy condition and water use}\label{prefire-canopy}

Field-level NDMI and NDVI were summarized from the pre-fire Sentinel-2 composite at 20~m and joined to OpenET and SSURGO AWC. Pearson and Spearman correlations described bivariate association. Ordinary least squares was used only as a descriptive line through field-level observations. Because ET integrates canopy development, climate, and management, it was interpreted as seasonal crop water use and never as metered irrigation. The field-scale design follows the broader need for explicit spatial support and uncertainty in agricultural remote sensing \citep{kalua2020vineyardet,narimani2026c}. Figure~\ref{fig:water} presents the primary relationships.

\subsection{Landscape spectral contrast}\label{landscape-contrast}

Within the perimeter, Sentinel-2 dNBR was summarized separately for vineyard and wildland pixels. Means, medians, and sample counts describe the magnitude of the contrast. The previously calculated pixel-level Welch statistic is retained in Table~\ref{tab:robustness} only as a naive diagnostic and is not used for inferential claims because adjacent 20~m pixels are spatially autocorrelated. Formal local inference instead relies on boundary segments with clustered uncertainty, and the conditional landscape model is interpreted descriptively.

The Landsat check compared area-wide dNBR magnitude, not the vineyard-wildland contrast. It therefore supports only the conclusion that the overall low landscape-average response was not unique to Sentinel-2; it does not independently validate a land-cover-specific effect.

\subsection{Vineyard-wildland boundary analysis}\label{boundary-analysis}

Two thousand vineyard-wildland boundary segments were sampled with signed distance $d$, defined as negative on the wildland side and positive on the vineyard side. At each bandwidth $h$, a local-linear model estimated the intercept difference at the boundary with standard errors clustered by segment.
\begin{equation}
y_{js}=\alpha+\tau V_{js}+\beta_{1}d_{js}+\beta_{2}(V_{js}\times d_{js})+u_{s}+\varepsilon_{js}
\label{eq:boundary}
\end{equation}
where $j$ indexes observations and $s$ indexes boundary segments; $y$ is dNBR; $V$ indicates the vineyard side; $d$ is signed distance to the boundary; $\tau$ is the intercept contrast; $u$ is the segment term; and $\varepsilon$ is residual error.

\subsection{Conditional landscape model}\label{conditional-model}

A 250~m grid inside the perimeter ($n = 5{,}097$ cells) supported a penalized spatial generalized additive model. Mean dNBR was related to vineyard fraction, elevation, slope, and mean growing-season ET, with separate smooths of normalized UTM easting and northing. Explained deviance and RMSE described fit, while $k$-nearest-neighbor Moran's~$I$ assessed residual spatial structure. Because residual autocorrelation remained substantial, the coefficients and nominal intervals are reported as conditional diagnostics rather than causal or fully spatially corrected effects. The within-vineyard association, descriptive land-cover contrast, and boundary balance are evaluated together in the Results.
\begin{multline}
\mathrm{dNBR}_{i}=\beta_{0}+\beta_{1}\mathrm{VF}_{i}+\beta_{2}\mathrm{ELEV}_{i}\\
\quad+\beta_{3}\mathrm{SLOPE}_{i}+\beta_{4}\mathrm{ET}_{i}+f_{x}(x_{i})+f_{y}(y_{i})+\varepsilon_{i}
\label{eq:gam}
\end{multline}
where $i$ indexes 250~m grid cells; VF is vineyard fraction; ELEV is elevation; SLOPE is terrain slope; ET is mean growing-season evapotranspiration; $f_{x}$ and $f_{y}$ are penalized coordinate smooths; and $\varepsilon$ is residual error.

\subsection{Potential smoke exposure}\label{smoke-methods}

Daily NOAA/NESDIS HMS smoke polygons from 23 October to 10 November 2019 were intersected with vineyard fields. A field-day was counted when any portion of a field intersected a smoke polygon; vineyard hectare-smoke-days summed the intersected vineyard area over all days. The measure indicates potential overhead smoke and is not a ground-level concentration, inhalation dose, or smoke-taint diagnosis. The daily footprint is evaluated together with post-fire greenness in the Results.
\begin{equation}
S_{j}=\sum_{t} \mathbf{1}\left(A_{j}\cap P_{t}\neq\emptyset\right)
\label{eq:smoke-days}
\end{equation}
where $S$ is the number of potential smoke-days for vineyard field $j$; $A$ is the field polygon; $P_{t}$ is the HMS smoke-polygon set on day $t$; and $\mathbf{1}(\cdot)$ is an indicator function.
\begin{equation}
\mathrm{HSD}=\sum_{t} \mathrm{area}\left(V\cap P_{t}\right)
\label{eq:hsd}
\end{equation}
where HSD is vineyard hectare-smoke-days, $V$ is the union of vineyard polygons, $P_{t}$ is the HMS smoke-polygon set on day $t$, and $\mathrm{area}(\cdot)$ returns the intersected vineyard area.

\subsection{Road access and network topology}\label{road-methods}

A drive network was reconstructed from an OpenStreetMap snapshot fixed to 23 October 2019. Vineyard centroids were linked to the nearest network node, nearest dead end, and nearest major road. Network size, directed edges, road length, intersection count, and dead-end fraction characterized structural accessibility. No traffic counts, capacities, travel times, or evacuation trajectories were used; the analysis therefore addresses topology and potential access rather than realized evacuation performance. The resulting access and topology metrics are summarized in the Results.

\subsection{Post-fire greenness}\label{recovery-methods}

April--September Sentinel-2 NDVI composites were summarized for 2018--2021. Fields were grouped by whether their centroid fell inside the fire perimeter. The 2019 season served as the immediate pre-fire baseline because the October ignition followed the growing-season composite. For field $i$ and post-fire year $t$, the recovery ratio was defined relative to 2019. Group means and Welch tests were retained as unadjusted field-level comparisons. Similar 2019 baseline NDVI provides some design reassurance, but the contrast remains observational and potentially influenced by spatial location, management, and the regional 2020--2021 drought. The trajectories and inside-outside differences are evaluated in the Results.
\begin{equation}
R_{it}=\frac{\mathrm{NDVI}_{it}}{\mathrm{NDVI}_{i,2019}}
\label{eq:recovery}
\end{equation}
where $R$ is the greenness recovery ratio for field $i$ in post-fire year $t$, $\mathrm{NDVI}_{it}$ is the growing-season NDVI in year $t$, and $\mathrm{NDVI}_{i,2019}$ is the field-specific 2019 baseline.

\subsection{Three-dimensional structure feasibility}\label{structure-feasibility}

OpenTopography, USGS 3DEP, GEDI L2A, and ICESat-2 availability were inventoried to determine whether event-day vertical fuels could be supported. Fourteen airborne point-cloud datasets intersected the area between 2003 and 2024; a 2013 Sonoma acquisition and a nearby 2017 acquisition were confirmed, but no dedicated 2019 Sonoma airborne collection was found. GEDI provided 45 monthly images from April 2019 through 2022 and 87,601 quality-screened RH98 pixels within the perimeter. Because these footprints undersample small vineyard blocks and airborne vintages do not coincide with the fire, Figure~\ref{fig:structure} is interpreted as a data-readiness result, not a parcel-level fuel analysis.

\subsection{Robustness, reproducibility, and ethics}\label{robustness-ethics}

Boundary robustness included bandwidths of 30, 60, 100, 150, and 300~m; continuity checks for elevation, slope, northness, and eastness; donut specifications; and an interior placebo boundary. The full set is reported in Table~\ref{tab:robustness} rather than an appendix. The workflow is parameterized from configuration files, executed by a stage orchestrator, and records every manuscript number in a machine-readable result registry. Analysis-ready products are archived in Zenodo \citep{kincade2026data}, and the public replication code is available on GitHub \citep{kincade2026code}. No human participants, identifiable human data, or animal experiments were involved; analyses used public environmental and infrastructure data.

\vspace{-6pt}
\section{Results}\label{results}

\subsection{Event and data coverage}\label{event-coverage}

\begin{figure*}[!t]
\centering
\includegraphics[width=\textwidth]{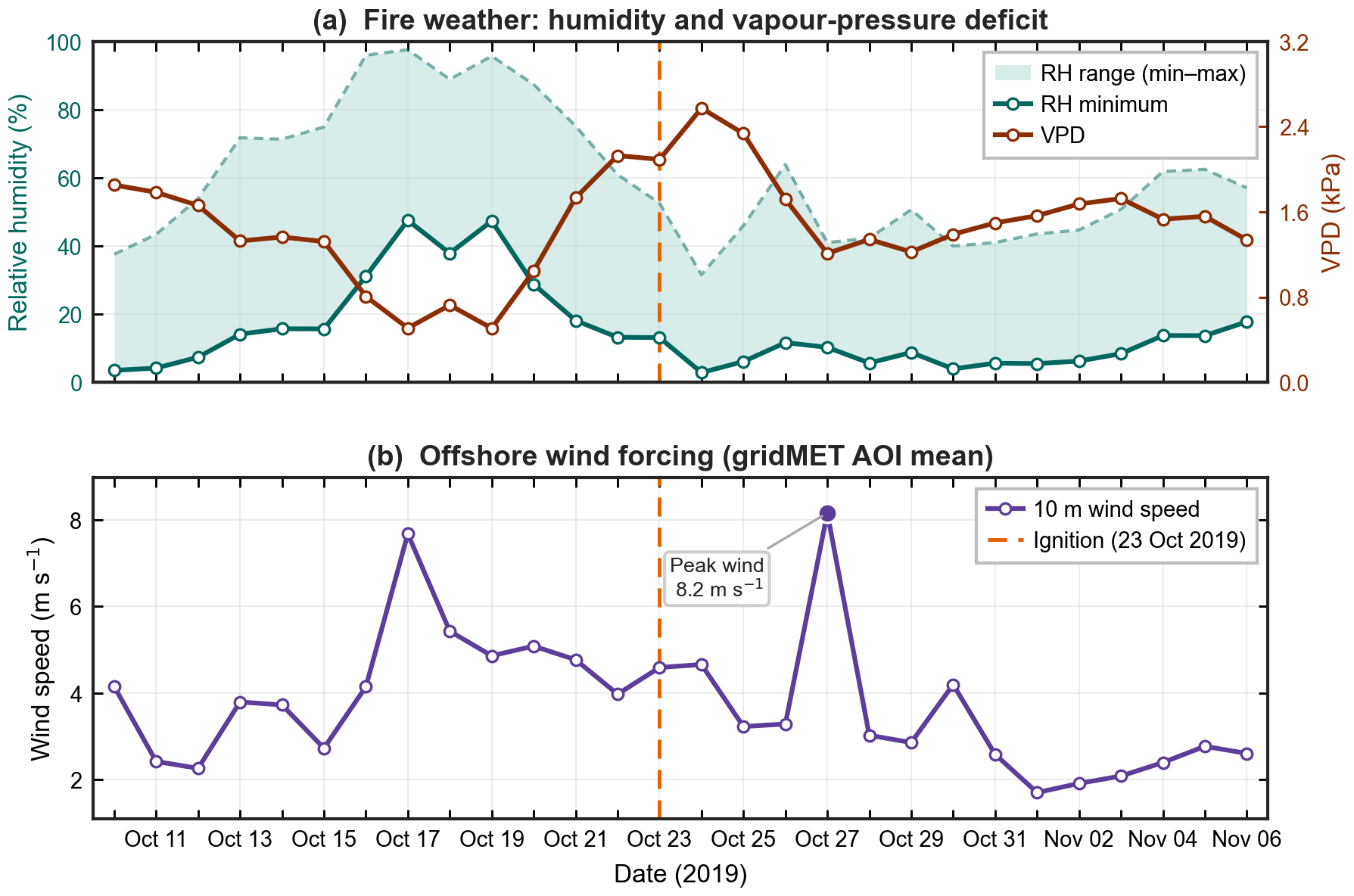}
\caption{Fire-weather and documented electrical-ignition context. (A) Area-mean daily relative humidity and vapor-pressure deficit from 10 October to 6 November 2019 (pre-ignition lead-in through containment). (B) Area-mean daily wind speed. The documented electrical-transmission ignition on 23 October is marked as event context. The onset occurred during exceptionally low humidity, elevated atmospheric demand, and an offshore wind episode; these fields are contextual and are not used to infer ignition causation.}
\label{fig:weather}
\end{figure*}

The open-data stack captured the principal stages of the event without treating every source as the same type of evidence. Figure~\ref{fig:studyarea} shows vineyards concentrated along the lower-relief western and southern portions of the fire landscape, adjacent to both surveyed structures and wildland vegetation. Figure~\ref{fig:chronology} demonstrates the sharp transition from pre-fire vegetation to an active smoke- and fire-obscured scene and then to the first clean post-fire spectral footprint. Figure~\ref{fig:weather} places the documented 23 October electrical ignition within an atmospheric sequence that included a minimum area-mean relative humidity of approximately 2.8\% on 24 October, elevated vapor-pressure deficit, and an area-mean wind maximum of approximately 8.2~m~s$^{-1}$ near 27 October. These coarse gridMET fields characterize event context rather than sub-field fire spread.

\subsection{Pre-fire canopy condition covaried with seasonal water use}\label{canopy-results}

\begin{figure*}[!t]
\centering
\includegraphics[width=\textwidth]{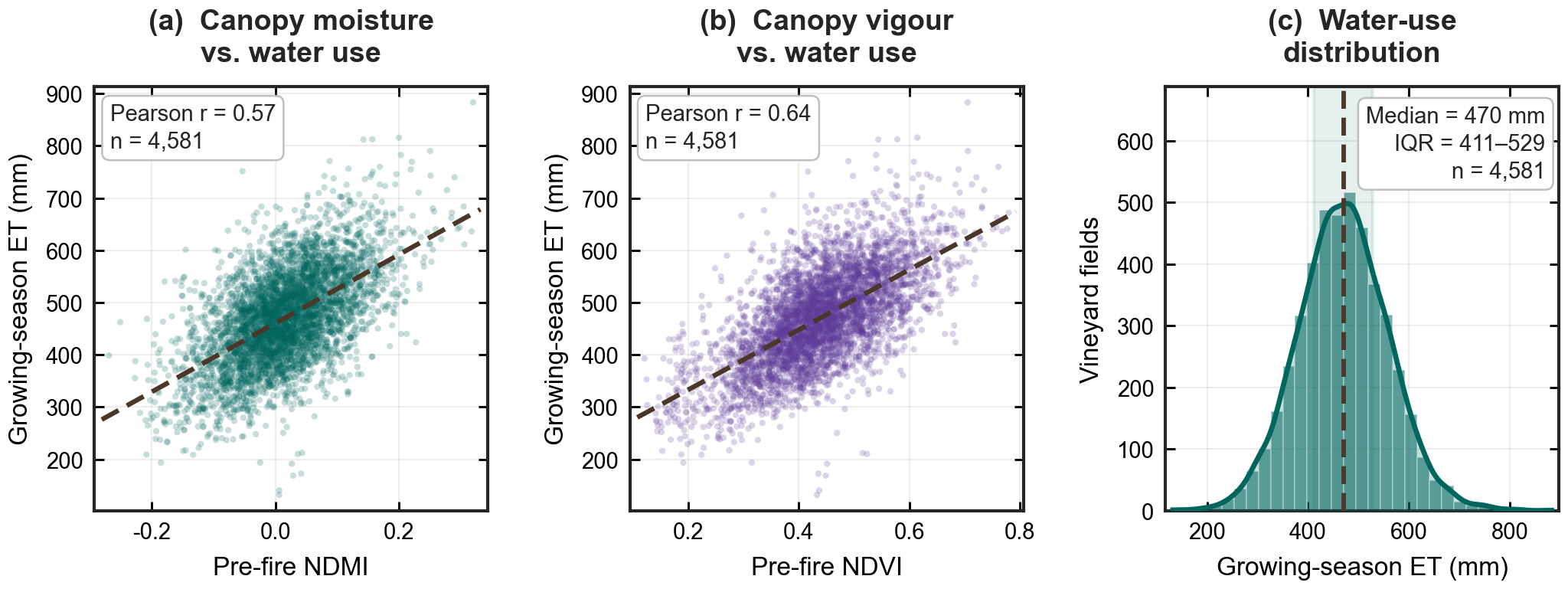}
\caption{Pre-fire vineyard canopy condition and seasonal water use. (A) Field-mean pre-fire Normalized Difference Moisture Index (NDMI) versus April--October OpenET evapotranspiration (ET) across 4,581 vineyard fields (Pearson $r = 0.567$). (B) Field-mean Normalized Difference Vegetation Index (NDVI) versus ET ($r = 0.639$). (C) Distribution of growing-season ET (mean 470.3~mm; median approximately 470~mm). ET is interpreted as remotely sensed crop water use, not metered irrigation.}
\label{fig:water}
\end{figure*}

Across 4,581 vineyard fields, mean pre-fire NDMI was 0.015 (SD 0.077), mean NDVI was 0.439, and mean growing-season ET was 470.3~mm. NDMI and NDVI were positively associated with ET (Pearson $r = 0.567$ and $0.639$; Spearman $\rho = 0.539$ and $0.620$, respectively; Figure~\ref{fig:water}). In contrast, mapped AWC explained little field-to-field variation: AWC to 100~cm correlated with NDMI at $r = 0.028$ ($p = 0.060$), AWC to 150~cm at $r = 0.044$ ($p = 0.003$), and AWC was not meaningfully associated with ET ($r = -0.026$). The result supports a narrow conclusion: field-scale canopy condition cohered with seasonal water use, whereas SSURGO AWC alone did not explain the observed variation. It does not identify individual irrigation applications or prove that management dominated soil controls.

\subsection{Vineyards displayed lower descriptive spectral response than wildland}\label{spectral-results}

\begin{figure*}[!t]
\centering
\includegraphics[width=\textwidth]{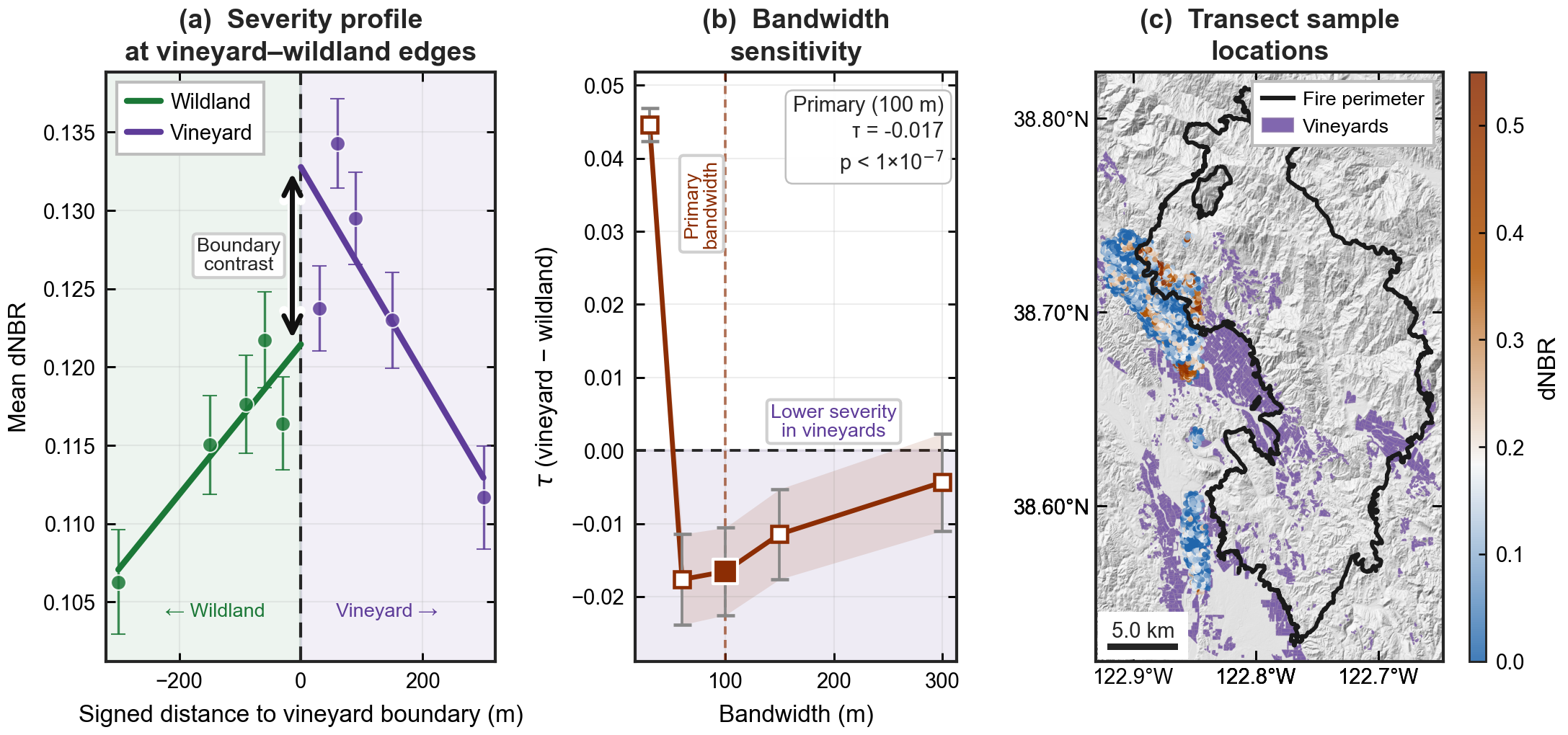}
\caption{Scale-dependent vineyard-wildland boundary contrast. (A) Binned mean dNBR with standard errors as a function of signed distance to the vineyard-wildland boundary (negative, wildland; positive, vineyard), with local linear fits. (B) Segment-clustered boundary contrasts across 30--300~m bandwidths; the primary 100~m estimate is $\tau = -0.0166$. (C) Spatial distribution of sampled boundary transects. Because slope continuity, donut, and placebo diagnostics fail, the pattern is reported as a spatial boundary association rather than a causal regression-discontinuity estimate.}
\label{fig:boundary}
\end{figure*}

Inside the perimeter, mean Sentinel-2 dNBR was 0.130 for 49,426 vineyard pixels and 0.337 for 746,849 wildland pixels; medians were 0.127 and 0.284. The mean contrast was $-0.207$ dNBR units, equivalent to a 61.4\% lower vineyard mean relative to wildland. Because pixels are not independent, this is a descriptive landscape contrast rather than a population-level significance test. The independent Landsat composite produced an area-wide mean dNBR of 0.086 compared with 0.082 for Sentinel-2, supporting the broad magnitude of landscape change but not independently reproducing the vineyard-wildland contrast. Figures~\ref{fig:boundary} and~\ref{fig:mechanisms} locate this strong unconditional pattern within the more cautious boundary and conditional analyses.

\subsection{The boundary signal was local, scale dependent, and noncausal}\label{boundary-results}

At the primary 100~m bandwidth, the local-linear boundary contrast was $-0.0166$ dNBR units (segment-clustered SE 0.0031). Estimates were $-0.0177$ at 60~m and $-0.0115$ at 150~m, but reversed to $+0.0446$ at 30~m and attenuated to $-0.0043$ at 300~m. More importantly, the identification diagnostics failed. Mean slope was 3.22 degrees on the vineyard side and 4.48 degrees on the wildland side (difference $-1.25$ degrees), whereas elevation and aspect components were approximately balanced. Excluding the immediate 100~m edge produced a null donut estimate ($+0.0068$; $p = 0.10$), and an interior placebo boundary generated a large wrong-signed estimate ($+0.0624$). Figure~\ref{fig:boundary} and Table~\ref{tab:robustness} therefore support only a scale-dependent spatial association near selected edges. They do not identify the causal effect of converting wildland to vineyard.

\subsection{Conditional modeling reversed the landscape contrast but left spatial structure}\label{gam-results}

The 250~m GAM explained 89.8\% of deviance with an RMSE of 0.055 dNBR units. Conditional on the coordinate smooths, elevation, slope, and ET, vineyard fraction had a positive coefficient of $+0.138$ and ET a positive coefficient of $+0.0076$. This sign reversal is scientifically useful but not definitive. Vineyards occupy lower, gentler terrain and differ spectrally from natural fuels; once location and covariates are conditioned on, the remaining vineyard fraction can represent a different comparison from the unconditional landscape contrast. Residual Moran's~$I$ was 0.519 (permutation $p = 0.001$), indicating that substantial spatial dependence remained. We therefore interpret the GAM as a diagnostic of confounding and spectral non-comparability, not as a causal severity model.

\begin{figure*}[!t]
\centering
\includegraphics[width=\textwidth]{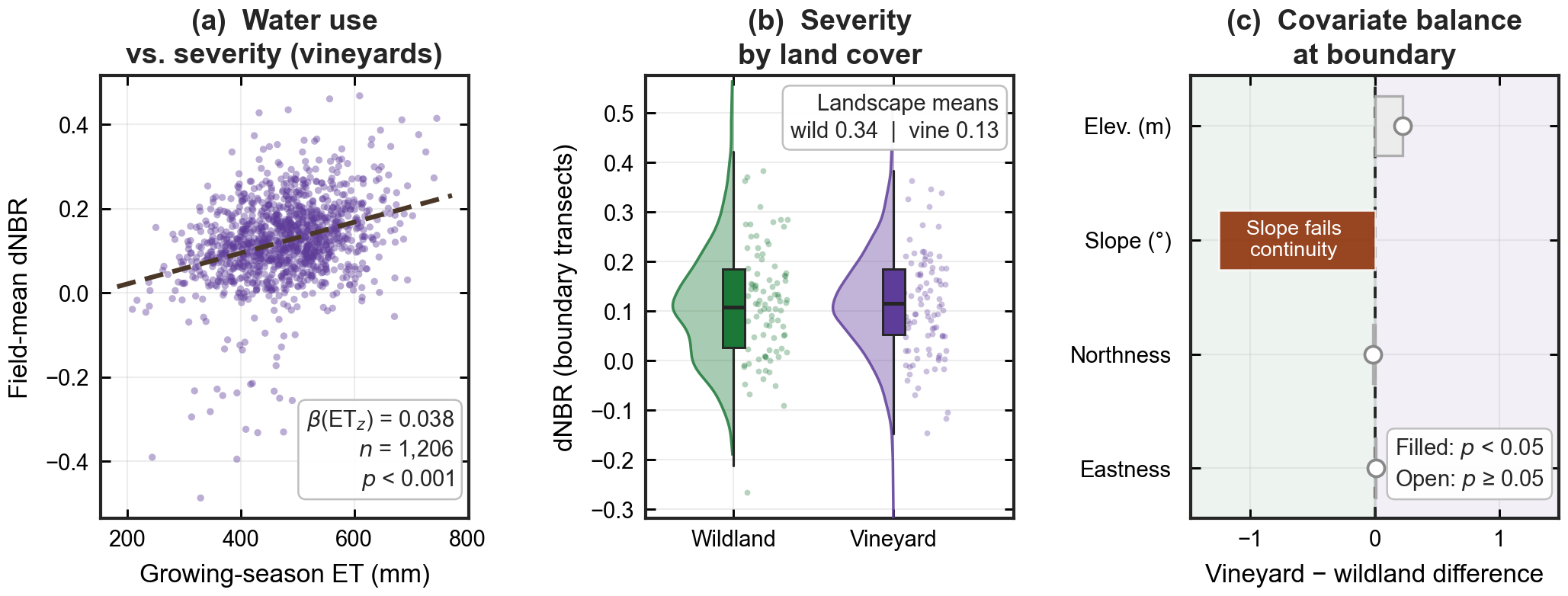}
\caption{Mechanisms and confounding in the spectral severity response. (A) Field-mean dNBR versus standardized growing-season ET among 1,206 vineyards with complete covariates. (B) Descriptive dNBR distributions for vineyard and wildland samples. (C) Covariate balance across the boundary: elevation and aspect are approximately continuous, whereas slope differs strongly. The positive ET coefficient is interpreted as covariance between pre-fire canopy development, water use, and a differenced spectral response, not as evidence that irrigation increases fire damage.}
\label{fig:mechanisms}
\end{figure*}

Within 1,206 vineyards with complete covariates, standardized ET was positively associated with field-mean dNBR ($\beta = +0.0384$, SE 0.0031). Elevation was also positive ($+0.0228$), whereas slope was small and negative ($-0.0024$). Figure~\ref{fig:mechanisms} shows why this does not imply that irrigation increased fire impact: fields with greater pre-fire canopy development and water use can exhibit a larger pre-post spectral difference even when their absolute fire effect is not greater.

\subsection{Road topology combined general access with rural fragility}\label{road-results}

\begin{figure*}[!t]
\centering
\includegraphics[width=\textwidth]{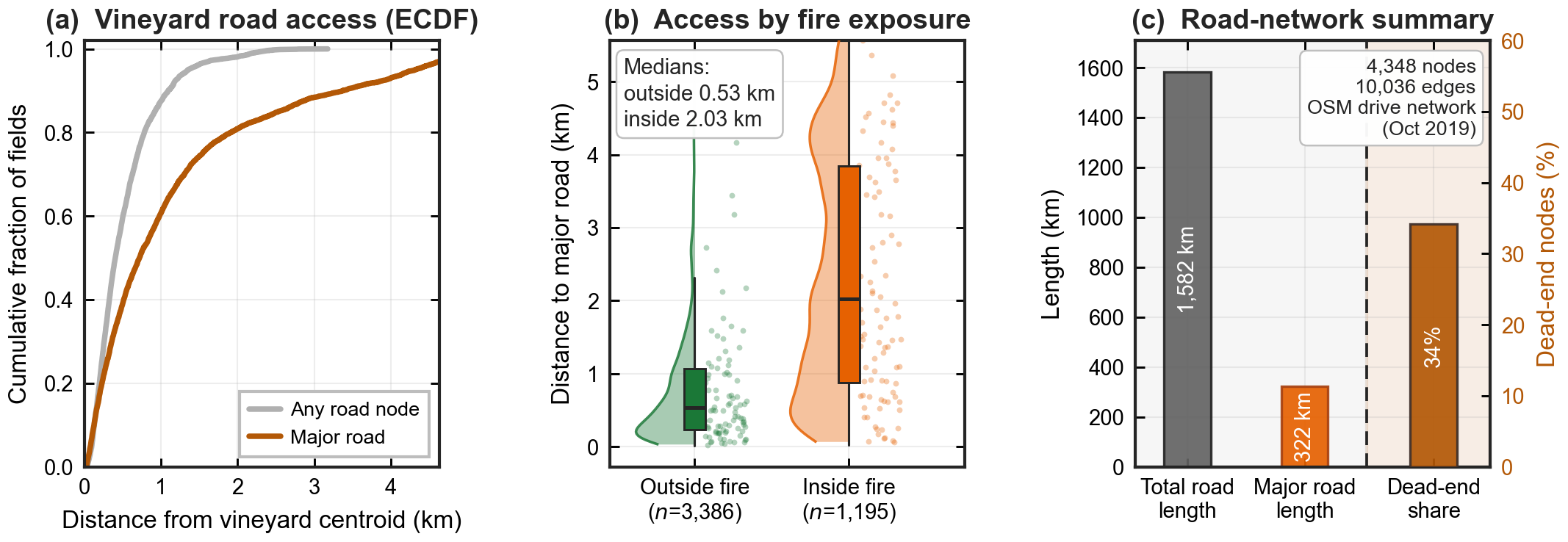}
\caption{Vineyard road access and network topology. (A) Cumulative distributions of vineyard distances to the nearest major road and to any road-network node. (B) Major-road distance for fields inside and outside the fire perimeter. (C) Summary of the ignition-date OpenStreetMap drive network: 4,348 nodes, 10,036 directed edges, 1,581.9~km of road, and a 34.2\% dead-end-node fraction. These metrics describe geometry and topology; traffic volume, capacity, and realized evacuation performance were not observed.}
\label{fig:roads}
\end{figure*}

The ignition-date road graph contained 4,348 nodes, 10,036 directed edges, and 1,581.9~km of road, of which 321.6~km were classified as major roads. The median vineyard centroid lay 404.4~m from the nearest road node, 591.6~m from the nearest dead end, and 719.5~m from a major road; the mean major-road distance was 1,204~m. At the same time, 1,487 nodes (34.2\%) were dead ends. Figure~\ref{fig:roads} therefore supports a dual interpretation: most fields were not remote from the network, but a substantial share of network endpoints indicates potential topological fragility. No conclusion about congestion or actual evacuation time is possible from these data.

\subsection{Smoke exposure was pervasive and spatially broader than direct burning}\label{smoke-results}

\begin{figure*}[!t]
\centering
\includegraphics[width=\textwidth]{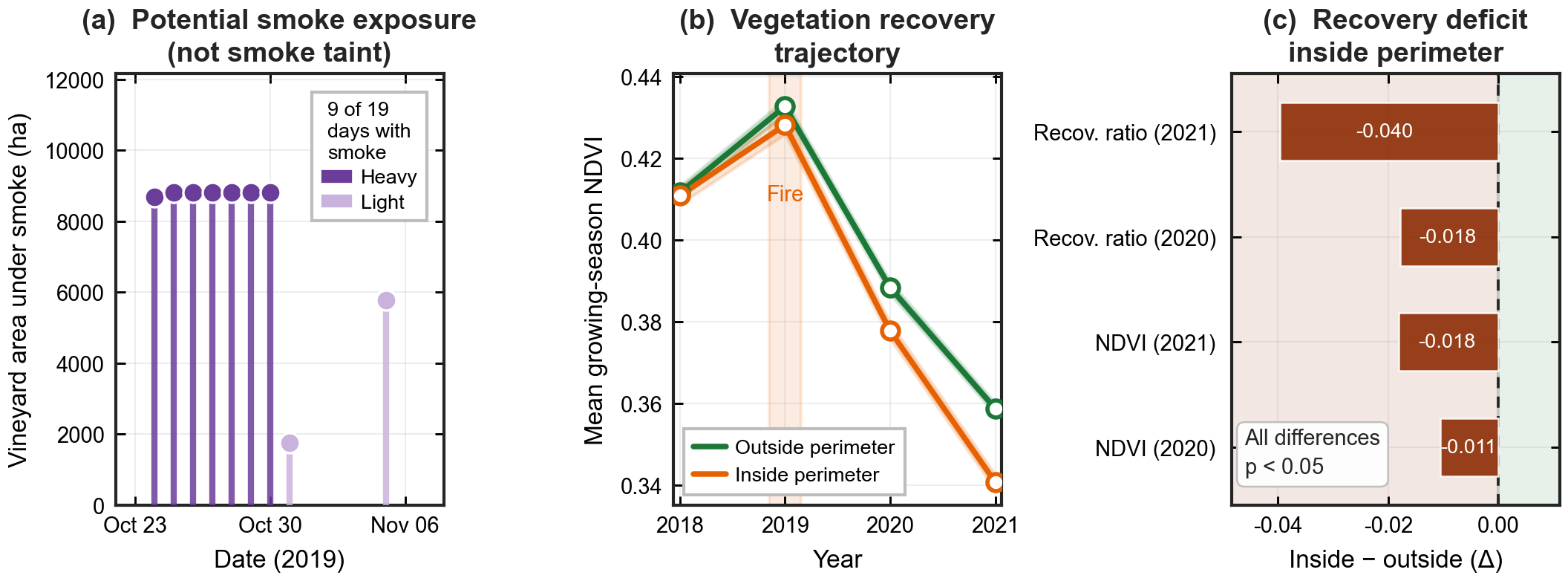}
\caption{Potential smoke exposure and post-fire vineyard greenness. (A) Daily vineyard area beneath NOAA/NESDIS Hazard Mapping System smoke polygons from 23 October to 10 November 2019. (B) Mean April--September NDVI trajectories for fields inside and outside the perimeter from 2018 to 2021. (C) Inside-minus-outside differences in annual NDVI and recovery ratios. Smoke polygons indicate potential overhead exposure, not ground-level PM$_{2.5}$ or grape smoke taint; NDVI measures greenness, not yield or wine quality.}
\label{fig:smoke}
\end{figure*}

HMS smoke overlapped the study area on 9 of 19 days. All 4,581 vineyard fields intersected a smoke polygon on at least one day, with a mean of 7.78 potential smoke-days per field and a maximum of 9. Summed across days, the footprint totaled 69,101.6 vineyard hectare-smoke-days. Figure~\ref{fig:smoke}A shows that most vineyard area was beneath heavy mapped smoke from 24 through 30 October, followed by smaller light-smoke footprints on 31 October and 5 November. These values describe potential overhead exposure only.

\subsection{Inside-perimeter vineyards retained a greenness deficit through 2021}\label{recovery-results}

The 2019 pre-fire growing-season baseline was similar for inside- and outside-perimeter fields (mean NDVI 0.4282 and 0.4327; difference $-0.0046$; $p = 0.11$). In 2020 the corresponding means were 0.3778 and 0.3883, and in 2021 they were 0.3407 and 0.3587. Recovery ratios declined to 0.8918 inside versus 0.9096 outside in 2020 and 0.8145 versus 0.8543 in 2021. The inside-minus-outside ratio difference widened from $-0.0178$ to $-0.0397$. Figure~\ref{fig:smoke}B--C thus shows a persistent, increasing greenness deficit relative to a similar 2019 baseline. Both groups declined during the regional drought period, so the result is an unadjusted differential trajectory rather than proof that the fire alone caused the full decline.

\subsection{Structural data readiness remained a limiting component}\label{structure-results}

\begin{figure*}[!t]
\centering
\includegraphics[width=\textwidth]{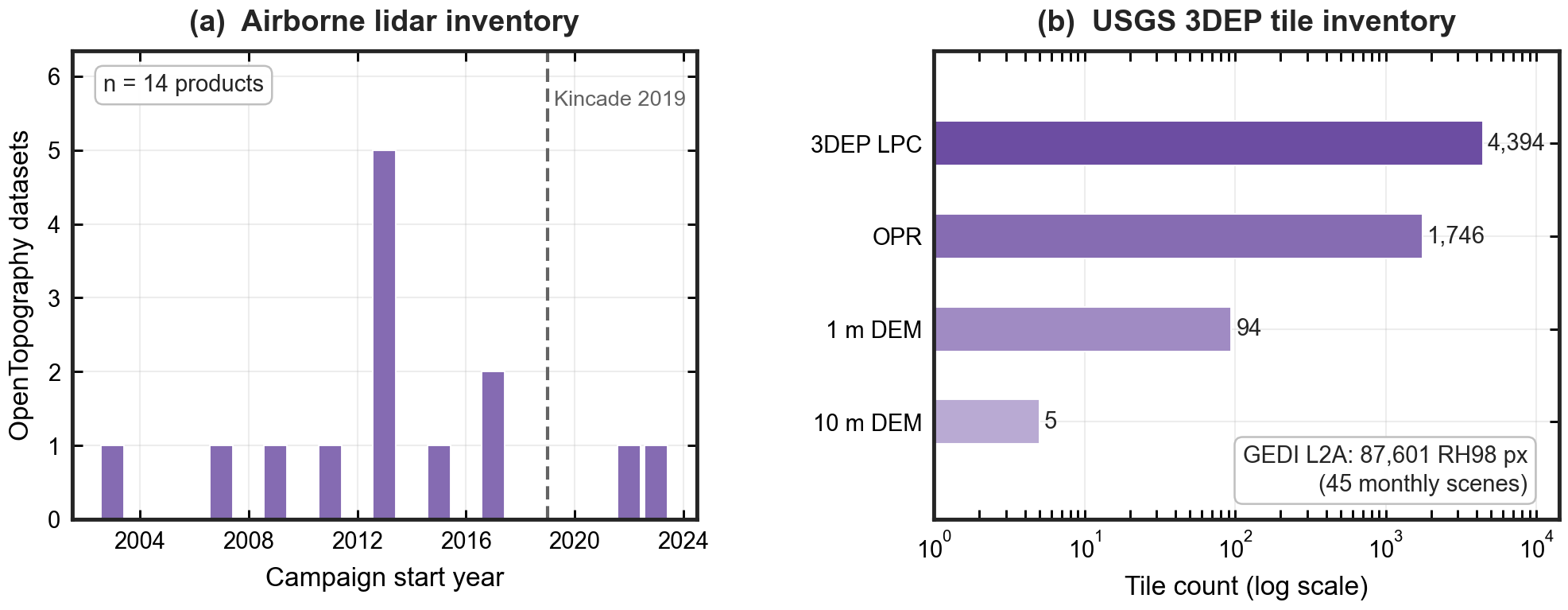}
\caption{Three-dimensional structure data availability. (A) Airborne lidar campaigns intersecting the study area by acquisition year (14 OpenTopography datasets spanning 2003--2024). (B) USGS 3DEP product counts and GEDI L2A coverage (45 monthly images and 87,601 quality-screened RH98 pixels within the perimeter). No dedicated 2019 Sonoma airborne campaign was found; these data therefore document structural baseline and landscape context rather than event-day fuel loads.}
\label{fig:structure}
\end{figure*}

Figure~\ref{fig:structure} documents extensive three-dimensional holdings but no temporally matched event-day airborne survey. The 2013 Sonoma acquisition can characterize historical structure, and GEDI can characterize portions of the surrounding woody landscape, but neither supports defensible parcel-level 2019 vineyard fuel loads. This negative result is operationally important: routine pre-fire lidar over working landscapes would be required to distinguish canopy height, inter-row fuels, and ladder-fuel continuity at the scale relevant to agricultural fire planning.

\subsection{A multidimensional evidence synthesis}\label{synthesis}

\begin{figure*}[!t]
\centering
\includegraphics[width=\textwidth]{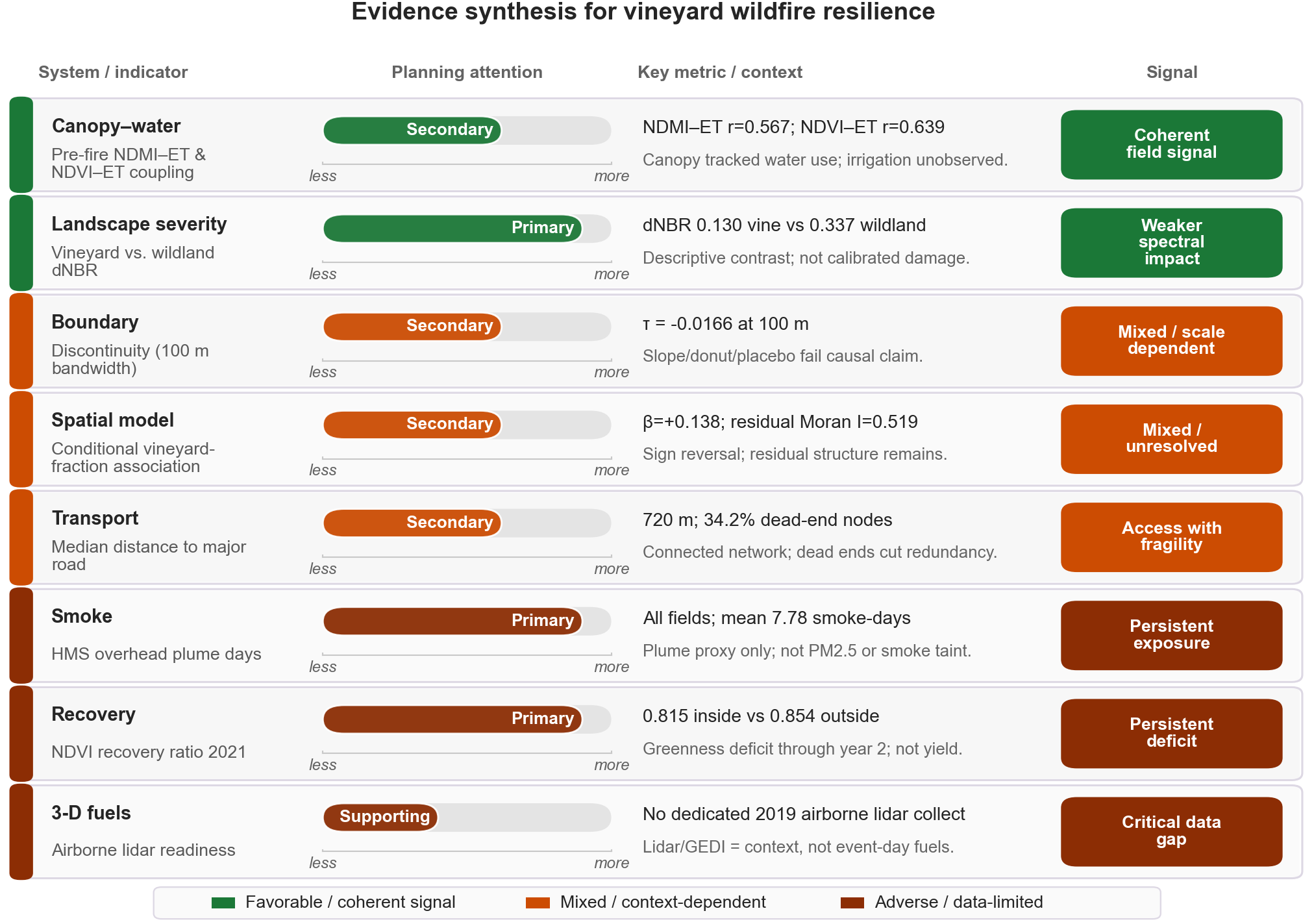}
\caption{Evidence synthesis for vineyard wildfire resilience. Eight system indicators are organized by planning attention (primary, secondary, or supporting), with the corresponding metric/context and qualitative signal. Landscape severity, smoke exposure, and recovery are treated as primary planning evidence; canopy-water status, boundary behavior, the conditional spatial model, and transport access provide secondary evidence; and three-dimensional fuel structure is supporting because no dedicated 2019 airborne lidar campaign was available. Planning attention is not a statistical-confidence ranking, and the synthesis intentionally preserves mixed, adverse, and data-limited signals rather than collapsing them into a composite resilience score.}
\label{fig:synthesis}
\end{figure*}

Figure~\ref{fig:synthesis} summarizes eight evidence domains without a composite resilience score. Planning-attention labels organize decision relevance rather than statistical certainty. Landscape severity, smoke, and recovery are primary; canopy-water status, boundary behavior, the conditional spatial model, and transport are secondary; and 3-D fuels are supporting because no dedicated 2019 airborne lidar campaign was available. The corresponding signals remain deliberately heterogeneous: coherent canopy-water coupling and weaker vineyard spectral impact coexist with scale-dependent boundary behavior, unresolved residual spatial structure, access fragility, persistent smoke exposure, a recovery deficit, and a critical data gap.

\begin{table*}[!t]
\centering
\caption{Principal quantitative findings.}
\label{tab:findings}
\footnotesize
\begin{tabular}{@{}p{0.22\textwidth}p{0.20\textwidth}p{0.24\textwidth}p{0.26\textwidth}@{}}
\toprule
Result & Estimate & Context / uncertainty & Interpretation \\
\midrule
Vineyard inventory & 4,581 fields; 8,813.2~ha & 5~km download AOI & Regional agricultural exposure frame. \\
Analysis-area vineyard inventory & 2,622 fields; 5,798.3~ha & Perimeter + 2~km & Severity and boundary analysis frame. \\
Pre-fire NDMI and NDVI & 0.015 and 0.439 mean & NDMI SD 0.077 & Field canopy condition before the October fire. \\
Growing-season ET & 470.3~mm mean & Median approximately 470~mm & Seasonal crop water use, not applied irrigation. \\
NDMI--ET / NDVI--ET & $r = 0.567$ / $0.639$ & Spearman $\rho = 0.539$ / $0.620$ & Canopy condition covaried with water use. \\
Vineyard / wildland dNBR & 0.130 / 0.337 mean & Medians 0.127 / 0.284 & Vineyard mean 61.4\% lower; descriptive landscape contrast. \\
Boundary contrast at 100~m & $\tau = -0.0166$ & Cluster SE 0.0031 & Small local association; causal assumptions fail. \\
Conditional GAM & Deviance explained $= 89.8\%$; RMSE $= 0.055$ & $n = 5{,}097$ cells & High fit but incomplete spatial correction. \\
GAM vineyard-fraction coefficient & $+0.138$ & Nominal 95\% CI 0.128--0.149 & Conditional sign reversal; not causal. \\
GAM residual Moran $I$ & 0.519 & Permutation $p = 0.001$ & Substantial unmodeled spatial dependence. \\
Within-vineyard ET coefficient & $+0.0384$ per SD & SE 0.0031; $n = 1{,}206$ & Positive spectral/biomass association. \\
Potential smoke exposure & 9 of 19 days; mean 7.78 field-days & All 4,581 fields $\ge$1 day & Pervasive overhead smoke footprint. \\
Vineyard hectare-smoke-days & 69,101.6 ha-days & HMS polygon intersections & Spatial-temporal exposure proxy. \\
Road topology & 34.2\% dead-end nodes & 4,348 nodes; 10,036 edges & Potential rural network fragility. \\
Major-road access & 719.5~m median & Mean 1,204~m & General proximity but heterogeneous access. \\
2021 recovery ratio & 0.8145 inside / 0.8543 outside & Difference $-0.0397$; $p = 2.3\times10^{-10}$ & Persistent unadjusted greenness deficit. \\
3-D data readiness & No dedicated 2019 airborne collect & GEDI 45 monthly images; 87,601 RH98 pixels & Structural context, not event-day vineyard fuels. \\
\bottomrule
\end{tabular}
\end{table*}

\begin{table*}[!t]
\centering
\caption{Robustness diagnostics and their implications for interpretation.}
\label{tab:robustness}
\footnotesize
\begin{tabular}{@{}p{0.16\textwidth}p{0.22\textwidth}p{0.26\textwidth}p{0.28\textwidth}@{}}
\toprule
Analysis & Specification / diagnostic & Result & Interpretive implication \\
\midrule
Pixel contrast & Welch test on 20~m pixels & $t = -109.5$; $p$ below numerical precision & Not used for inference because pixels are spatially autocorrelated. \\
Landsat check & Area-wide dNBR mean & 0.086 Landsat vs 0.082 Sentinel-2 & Supports landscape magnitude only, not land-cover-specific replication. \\
Boundary bandwidth & 30~m & $\tau = +0.0446$ & Sign reversal at narrowest scale. \\
Boundary bandwidth & 60 / 100 / 150~m & $\tau = -0.0177$ / $-0.0166$ / $-0.0115$ & Small negative association over intermediate scales. \\
Boundary bandwidth & 300~m & $\tau = -0.0043$; $p = 0.20$ & Attenuates to null at broad scale. \\
Continuity & Elevation / northness / eastness & Approximately balanced & Some local design support. \\
Continuity & Slope & 3.22 vs 4.48 degrees; difference $-1.25$ & Key confounder discontinuity; causal RDD rejected. \\
Donut & Exclude immediate $\pm$100~m & $\tau = +0.0068$; $p = 0.10$ & Primary edge contrast disappears. \\
Donut & Exclude immediate $\pm$150~m & $\tau = +0.0102$; $p = 0.013$ & Wrong-signed positive residual contrast. \\
Placebo boundary & 100~m interior offset & $\tau = +0.0624$ & Large non-null background gradient. \\
Spatial GAM & Residual Moran $I$ & 0.519; $p_{\mathrm{sim}} = 0.001$ & Nominal coefficient SEs are optimistic. \\
Water-status interpretation & AWC correlations & NDMI $r = 0.028$ / $0.044$; ET $r = -0.026$ & Mapped soil AWC alone explains little; management dominance not inferred. \\
Recovery baseline & 2019 inside vs outside & 0.4282 vs 0.4327; $p = 0.11$ & Comparable baseline, but post-fire comparison remains observational. \\
Recovery trend & 2020 / 2021 ratio gap & $-0.0178$ / $-0.0397$ & Inside deficit widens during a regional drought period. \\
Smoke construct & HMS daily polygons & Overhead plume overlap only & No PM$_{2.5}$ concentration, dose, or smoke-taint claim. \\
Transport construct & Historical OSM graph & No traffic, capacity, or closure data & No realized evacuation-performance claim. \\
Lidar timing & Airborne inventory & No dedicated 2019 Sonoma collect & No event-day vertical-fuel claim. \\
\bottomrule
\end{tabular}
\end{table*}

\section{Discussion}\label{discussion}

\subsection{Lower spectral burn response did not equal complete resilience}\label{disc-spectral}

The most visible result is also the easiest to overstate: vineyard pixels had a much lower mean dNBR than surrounding wildland. At landscape scale, this is consistent with vineyards interrupting continuous wildland fuels through shorter trained canopies, inter-row gaps, irrigation-related water use, and valley-bottom placement. Yet dNBR was developed and calibrated primarily for natural vegetation, and managed canopies have different pre-fire biomass, phenology, soil exposure, and row geometry \citep{key2006severity,miller2007rdnbr}. A lower differenced spectral response can therefore reflect lower combustion, lower pre-fire biomass, rapid soil-background exposure, or combinations of these processes. The result is real as a sensor-observed contrast, but its mechanism is not uniquely identified.

The broader evidence shows why resilience cannot be reduced to that one contrast. Every field experienced potential overhead smoke, road access coexisted with many dead ends, and inside-perimeter greenness remained lower through the second post-fire growing season. The paper's central conclusion is consequently not that vineyards are fireproof. It is that immediate spectral impact, atmospheric exposure, network accessibility, structural-data readiness, and recovery describe different dimensions of the same event.

\subsection{Water use and canopy amount complicate interpretation}\label{disc-water}

Pre-fire NDMI and NDVI were coherently related to OpenET, whereas SSURGO AWC was weakly related to either canopy condition or ET. This pattern is compatible with field-specific differences in canopy development, water use, soil depth not captured by map-unit averages, microclimate, cultivar, phenology, and management. It does not permit a claim that irrigation caused the observed canopy state because applied water and management records were unavailable. The distinction matters for both wildfire and agricultural interpretation: OpenET measures evapotranspiration, not irrigation delivery.

The positive within-vineyard ET coefficient on dNBR provides another caution. More developed or wetter canopies may have a larger pre-fire NBR and therefore a larger differenced response after disturbance. That spectral arithmetic can produce a positive ET--dNBR association without implying that water application increases flammability. Leaf-scale work in almond orchards has combined PROSPECT-PRO radiative transfer with Gaussian process regression and evaluated nitrogen estimates against independent leaf measurements \citep{chakraborty2025almondn}. Adapting that measurement-to-trait approach to fire-affected grapevines could help distinguish physiological recovery from changes in canopy cover. Future work should combine field-calibrated combustion or damage observations with ET, canopy fraction, cover-crop condition, and row-scale structure before assigning a management mechanism.

\subsection{Boundary analysis improved honesty more than certainty}\label{disc-boundary}

The boundary design was valuable because it converted a visually compelling map contrast into a falsifiable identification exercise. The result did not survive that exercise as a clean causal estimate. Slope changed across the boundary, the narrowest bandwidth reversed sign, the 100~m donut estimate was null, and the placebo boundary was strongly non-null. These diagnostics indicate that vineyard placement and background fire gradients structure the apparent edge effect.

The same principle applies to the GAM. Its high explained deviance could appear impressive, but the residual Moran's~$I$ of 0.519 shows that smooth coordinates did not fully account for spatial dependence. Nominal coefficient intervals are therefore too optimistic. Recent data-driven wildfire studies and spatially validated structure-loss work reinforce that flexible models must be evaluated against realistic spatial structure rather than random or model-internal fit alone \citep{biswas2025geoai,farajpoor2026palisades}. Here, the sign reversal is most useful as a warning: the unconditional landscape comparison and the conditional grid comparison answer different questions, and neither supports a universal firebreak claim.

\subsection{Smoke, roads, and recovery reveal cascading exposure}\label{disc-cascade}

The smoke analysis expands the spatial definition of impact. Direct burning affected a subset of vineyards, whereas overhead smoke reached every mapped field for at least one day. This does not establish ground-level concentration, fruit uptake, or sensory taint, but it identifies the scale and timing of potential exposure during an agriculturally important period. Chemical studies show that smoke taint depends on volatile phenols, glycosides, dose, timing, variety, and winemaking processes, none of which can be inferred from HMS polygons alone \citep{krstic2015smoketaint,summerson2021smokereview}. The correct operational use of these maps is therefore to prioritize monitoring and sampling, not to classify crop loss.

Road evidence is similarly bounded. Median distances indicate broad network access, but the high dead-end fraction points to potential dependence on a limited set of links. Existing Kincade evacuation studies show that household decisions and network demand are dynamic \citep{xu2023kincade}, while behavioral and network-mobility studies show that realized evacuation depends on departure timing, warning, route, destination, and host-community conditions \citep{wong2023evacuee,borody2025evacuation}. The present topology can identify locations where redundancy deserves examination; it cannot reconstruct evacuation performance.

Recovery provides the clearest longer-duration contrast. Comparable 2019 baselines were followed by a widening inside-perimeter NDVI deficit through 2021. That trajectory is consistent with lingering fire-related effects, replanting or canopy-management changes, and differential drought sensitivity, but it remains unadjusted. Post-fire ecological responses documented elsewhere in the Kincade landscape likewise vary by species and habitat \citep{lumpkin2026birds}. A stronger next step would use matched fields, longer pre-fire histories, management records, and yield or vine-removal observations to separate fire effects from regional drought and operational decisions. In a different crop-stress setting, Sentinel-2 time-series analysis of broomrape-infested tomato fields used growing-degree-day alignment to compare canopy trajectories across phenological stages \citep{narimani2025broomrape}. A comparable phenology-aware design would strengthen future vineyard recovery comparisons; the present NDVI trajectories remain a greenness screen rather than a causal recovery attribution.

\subsection{Urban-rural informatics and decision use}\label{disc-decision}

The study fits an urban-rural informatics perspective because it organizes heterogeneous public records around decisions rather than around data availability. The map stack can support several practical actions: identify vineyard-wildland edges that need field inspection; direct smoke sampling to fields with repeated plume overlap; flag areas where road redundancy is limited; distinguish immediate spectral impact from longer recovery; and specify where event-day three-dimensional fuel data are absent. Open, updateable screening layers can complement municipal and regional inventories when their assumptions and spatial support are explicit \citep{narimani2026b}. Comparable cross-hazard resilience research has also used multidimensional, data-driven evidence to support planning, underscoring the value of transparent integration across system components \citep{pour2025ml}.

This decision use differs from a composite resilience index. Figure~\ref{fig:synthesis} deliberately separates planning attention from signal direction. Primary rows capture landscape spectral severity, smoke exposure, and recovery; secondary rows retain canopy-water status, boundary behavior, the conditional spatial model, and transport; and 3-D fuels are supporting because event-matched airborne lidar is unavailable. These labels do not rank statistical certainty. They preserve the fact that weaker immediate spectral impact can coexist with scale-dependent edge behavior, unresolved spatial confounding, transport fragility, persistent smoke exposure, and delayed greenness recovery. The structure is transferable to other working landscapes, including orchards, rangelands, and peri-urban agriculture, provided that crop phenology, fuel architecture, and local infrastructure are redefined rather than copied. Rural wildfire policy similarly benefits from separating observed system conditions from claims about realized protection \citep{steel2025cwpp}.

\subsection{Limitations}\label{limitations}

Several limitations define the boundary of the findings. First, dNBR is a spectral response, not a field-calibrated vineyard damage measure; Landsat confirms only the broad landscape magnitude. Second, vineyard and wildland pixels differ in terrain, location, biomass, and reflectance, and pixel-level samples are spatially autocorrelated. Third, the boundary design fails causal continuity and placebo diagnostics. Fourth, the GAM leaves substantial residual spatial autocorrelation, so its nominal intervals are optimistic. Fifth, OpenET is monthly crop water use rather than applied irrigation, SSURGO AWC is a map-unit estimate, gridMET is too coarse for row-scale fire weather, and HMS indicates overhead smoke rather than surface concentration or taint. Sixth, the road analysis contains no traffic, capacity, closure, or warning data. Seventh, recovery is an unadjusted greenness comparison through 2021, not a measure of vine survival, yield, fruit quality, or economic recovery. Finally, no dedicated 2019 airborne lidar acquisition was found, preventing event-day analysis of vineyard canopy and inter-row fuels.

These limitations also indicate the most valuable observations for future fire events: date-matched airborne lidar, field damage surveys for working crops, irrigation and canopy-management records, grape chemistry and ground PM$_{2.5}$, time-stamped closures and traffic, and longer recovery histories. Post-fire soil and hydrologic monitoring should likewise be connected to mapped severity and management units rather than treated as a separate environmental endpoint \citep{neris2023soilerosion,usgs2025postfiresoils}.

\subsection{Conclusion}\label{conclusion}

During the 2019 Kincade Fire, vineyards showed a markedly lower immediate spectral burn response than surrounding wildland, but that contrast weakened under local and conditional scrutiny and did not identify a universal firebreak effect. The same fields experienced widespread potential smoke exposure, a road network with substantial dead-end structure, and a persistent post-fire greenness deficit. The most defensible resilience interpretation is therefore multidimensional: managed agricultural landscapes can interrupt some forms of immediate fire impact while remaining vulnerable through atmospheric exposure, infrastructure dependence, and recovery. By placing every result beside its spatial unit, diagnostic test, and claim boundary, this open workflow provides a transparent basis for agricultural-interface wildfire planning without converting heterogeneous evidence into false certainty.

\FloatBarrier

\section*{Data availability statement}\label{data-availability-statement}

All public source products and collection identifiers are listed in Table~\ref{tab:datasets}, and persistent source links, retrieval metadata, processing roles, and checksums are documented in the repository data manifest. The analysis-ready vineyard-field tables, study boundaries, manuscript tables, model diagnostics, boundary-analysis outputs, smoke-exposure and recovery summaries, figure-source data, final figures, the result registry, manifests, and related derived products are openly available in Zenodo at \url{https://doi.org/10.5281/zenodo.22713181} \citep{kincade2026data}. The same record provides the clipped Sentinel-2, Landsat-8, OpenET, gridMET, HMS smoke, OpenStreetMap, terrain, and supporting vineyard and structure inputs used to rebuild the analyses, subject to the licenses and redistribution conditions of the original providers. Companion replication code is available on GitHub \citep{kincade2026code}.

\section*{Code availability statement}\label{code-availability-statement}

Replication code, configuration files, environment specifications, automated checks, and reproducibility documentation are openly available at \url{https://github.com/MohammadrezaNarimaniUCDavis/Kincade_Vineyard_Wildfire_Resilience} \citep{kincade2026code}. The repository includes the scripted workflow used to rebuild the principal tables, model diagnostics, boundary analyses, smoke and recovery summaries, and figures from the archived Zenodo data products \citep{kincade2026data}.

\section*{Ethics statement}\label{ethics-statement}

No human participants, identifiable human data, animal subjects, or biological specimens were involved; the study used public geospatial and environmental datasets.

\section*{Author contributions}\label{author-contributions}

Parastoo Farajpoor: Conceptualization, Methodology, Software, Formal analysis, Investigation, Data curation, Visualization, Writing -- original draft, Writing -- review and editing.
Mahla Ardebili Pour: Conceptualization, Methodology, Software, Formal analysis, Investigation, Data curation, Validation, Writing -- original draft, Writing -- review and editing.
Mohammad Bagher Ghiasi: Conceptualization, Methodology, Software, Formal analysis, Investigation, Data curation, Visualization, Writing -- original draft, Writing -- review and editing.
Mohammadreza Narimani: Conceptualization, Methodology, Software, Formal analysis, Investigation, Data curation, Visualization, Writing -- original draft, Writing -- review and editing, Project administration.

\section*{Funding}\label{funding}

The authors declare that no external financial support was received for the research, authorship, or publication of this article.

\section*{Conflict of interest}\label{conflict-of-interest}

The authors declare that the research was conducted in the absence of any commercial or financial relationships that could be construed as a potential conflict of interest.

\section*{Generative AI statement}\label{generative-ai-statement}

During the preparation of this work, the authors used ChatGPT to improve grammatical accuracy, refine sentence structure, and enhance visualizations. All AI-generated revisions were thoroughly reviewed and edited by the authors to ensure relevance and accuracy.

\section*{Acknowledgments}\label{acknowledgments}

The authors thank CAL FIRE, WFIGS/NIFC, the California Department of Water Resources, the U.S.\ Geological Survey, USDA NRCS, Copernicus, the Climatology Lab, OpenET, NOAA/NESDIS, NASA, OpenTopography, and OpenStreetMap contributors for maintaining the public data resources used in this study. This study contains modified Copernicus Sentinel data. OpenStreetMap data are available under the Open Database License.

\section*{Abbreviations}\label{abbreviations}

AWC, available water capacity; DINS, Damage Inspection; dNBR, differenced Normalized Burn Ratio; ET, evapotranspiration; GAM, generalized additive model; GEDI, Global Ecosystem Dynamics Investigation; GEE, Google Earth Engine; HMS, Hazard Mapping System; NDMI, Normalized Difference Moisture Index; NDVI, Normalized Difference Vegetation Index; NBR, Normalized Burn Ratio; OSM, OpenStreetMap; RdNBR, relativized differenced Normalized Burn Ratio; VPD, vapor-pressure deficit; WUI, wildland-urban interface.

\bibliographystyle{IEEEtranN}
\bibliography{references}

\end{document}